\documentclass[prb,nofootinbib,twocolumn,superscriptaddress]{revtex4} 
\usepackage{graphicx}% Include figure files
\usepackage{dcolumn}% Align table columns on decimal point
\usepackage{bm}% bold math 
\usepackage{threeparttable}
\usepackage{times}
\usepackage{mathptmx}
\usepackage{lscape}
\usepackage{natbib}
\usepackage{amsmath}
\usepackage{amssymb}
\usepackage{braket}
\usepackage{comment}
\usepackage{color}

\def\degree{\kern-.2em\r{}\kern-.3em}

\begin{document}

%\preprint{APS/123-QED} 

\title{ Microscopic Geometry Rules Ordering Tendency for Multicomponent Disordered Alloys   }

\author{Koretaka Yuge}
\affiliation{
Department of Materials Science and Engineering,  Kyoto University, Sakyo, Kyoto 606-8501, Japan\\
}%

\author{Shouno Ohta}
\affiliation{
Department of Materials Science and Engineering,  Kyoto University, Sakyo, Kyoto 606-8501, Japan\\
}%

\begin{abstract}
{ Short-range ordering (SRO) tendency for disordered alloys is considered as competition between chemical ordering and geometric (mainly, difference in atomic radius for constituents) effects. Especially for multicomponent (including the so-called high entropy alloys (HEAs) near equiatomic composition), it has been considered as difficult to systematically characterize the SRO tendency only by geometric effects, due mainly to the fact that (i) chemical effects typically plays significant role, (ii) near equiatomic composition, we cannnot classify which elements belong to solute or solvent, and (iii) underlying lattice for pure elements can typically differ from each other. Despite these facts, we here show that SRO tendency for seven fcc-based alloys including subsystems of Ni-based HEAs, can be well characterized by geometric effects, where corresponding atomic radius is defined based on atomic configuration with special fluctuation, measured from ideally random structure. The present findings strongly indicate the significant role of geometry in underlying lattice on SRO for multicomponent alloys.
  }
\end{abstract}

%\pacs{81.90.+c \sep 61.05.-a \sep 05.20.Gg \sep 05.10.-a \sep 02.30.Zz }

\maketitle

\section{Introduction}
%\begin{itemize}
%\item
For binary disordered alloys, there have been experimental and/or theoretical attempts to universally characterize short-range order (SRO) tendency based on the difference (or ratio) of constituent atomic radius. These are commonly based on the intuition that ordering tendency (i.e., unlike-atom pairs are preffered) is enhanced when constituent atomic radius exhibit significant differences due to reducing strain energy. 
%\item
This intuition fails, i.e., SRO tendency for binary alloys cannot be generally characterized only by the difference between constituent atomic radius, since chemical effects, coming from many-body interactions in alloys, play central role on SRO.\cite{sro-j} Following theoretical investigations quantitatively point out the significant chemical effect on SRO,\cite{chem1, chem2} which naturally makes the trends that the relationship between SRO for binary alloys and geometric effects (including differences in atomic radius) have not been focused on so far. 
%\item
Despite these facts, we recently reveal\cite{em-sro} that although trends in SRO for fcc-based binary alloys cannot be characterized by conventional Goldschmidt or DFT-based atomic radius even qualitatively, those are well-characterized by effective atomic radius of specially-selected microscopic structure derived only from information of underlying geometry (i.e., lattice). The results strongly indicate that ratio of constituent atomic radius in short-range ordered structure covariance deviation, w.r.t. configurational density of states (CDOS) for \textit{non-interacting} system, from linear average for constituent pure elements  can characterize the SRO tendency. 
%\item
Although recent DFT-based theoretical investigations amply address SRO tendencies not only for binary, but also for multicomponent alloys at or near equiatomic composition such as CoCrNi and CoFeNi for ternary, and Mo- and Ni-based quaternary and quinary alloys due to the recently attractive attention to the so-called high entropy alloys (HEA) that can exhibit super-high strength compared to conventional alloys, very little has been focused on SRO in multicomponent (three or more components) systems in terms of geometric contributions, due mainly to the fact that (i) near equiatomic composition, we cannot clearly classify whether the selected element is solute or solvent, (ii) underlying lattice for pure constituent can differ from that for HEA and (iii) it has been believed that chemical effects other than geometric effects play central role on SRO for binary systems. 
%\item
Here we extend our previous approach on binary system to systematically characterizing ordering tendency for seven equiatomic ternary alloys, by explicitly employing non-conventional coordinates (i.e., not conventional in generalized Ising systems) along chosen pair probability, where quantitative formulation for relationship between conventional and non-conventional coordinates for multicomponent discrete systems under \textit{constant composition} has recently been clarified by our study. 
%\item
We find that SRO tendencies in terms of neighboring pair probability can be universally, well characterized by ratio of atomic radius between of specially-fluctuated ordered structure and of ideally random structure, which is discussed in detail below. 
%\end{itemize}

\section{Methodology}
%\begin{itemize}
%\item
In order to quantitatively represent any atomic configuration for multicomponent systems, we here employ generalized Ising model\cite{ce} (GIM) that provides complete orthonormal basis functions enables to exactly  describe physical quantities as a function of atomic configuration for classical discrete systems under constant composition. In the present A-B-C ternary systems, we define spin variables $\sigma_i$ to specify atomic occupation of A, B or C at lattice point $i$ as $\sigma_i=+1$ for A, $\sigma_i=0$ for B and $\sigma_i=-1$ for C element. Under this definition, basis functions at a single lattice point are given by
\begin{eqnarray}
\label{eq:basis}
\phi_0 = 1,\quad \phi_1 =\sqrt{\frac{3}{2}}\sigma_i,\quad \phi_2 = \sqrt{2}\left(\frac{3}{2}\sigma_1^2 - 1\right),
\end{eqnarray}
and higher-order multisite correlations $\Phi$ can be obtained by taking average of corresponding products of the above basis functions over all lattice points. 
%\item
Once we construct complete set of basis, we recently find that for any number of components, canonical average of multisite correlation function can be given by
\begin{eqnarray}
\Braket{\Phi_r}\left(T\right) \simeq \Braket{\Phi_r}_1 - \sqrt{\frac{\pi}{2}} \Braket{\Phi_r}_2\frac{U_r - U_0  }{k_{\textrm{B}}T},
\end{eqnarray}
where $\Braket{\quad}_1$ and $\Braket{\quad}_2$ respectively denotes taking arithmetic average and standard deviation over all possible atomic configurations for \textit{non-interacting} system.
$U_r$ and $U_0$ represents potential energy of the so-called "projection state (PS)"\cite{em2} and special quasirandom structure (SQS, that mimic perfect random state).\cite{sqs}
Since we have shown that structures of PS and SQS, and $\Braket{\quad}_1$ and $\Braket{\quad}_2$ can be known from system \textit{before} applying many-body interactions, we can \textit{a priori} know their information without any thermodynamic information. 
This directly means that once we obtain the structure of PS and SQS, we can systematically characterize temperature dependence of ordering tendency for multicomponent alloys.
%\item
However, for multicomponent system, basis functions in Eq.~\eqref{eq:basis} does not provide intuitive interpretation of which like- and/or unlike-atom pairs are preferred and/or disfavored, which is contrary to the case of binary system. For multicomponent systems, practical problemes are (i) pair probabilities are obtained by linear combination of the basis functions, leading to accumulating predictive error and (ii) a set of pair probability does not form orthonormal basis. To overcome these problems, we recently modify Eq.~\eqref{eq:basis} that can be applied to canonical average of any linear combination of orthonormal basis functions. For pair probability of $Y_{IJ}$, this is given by
\begin{widetext}
\begin{eqnarray}
\label{eq:y}
\Braket{Y_{IJ}}\left(T\right) \simeq \Braket{Y_{IJ}}_1 - \sqrt{\frac{\pi}{2}} \frac{\Braket{Y_{IJ}}_2}{k_{\textrm{B}}T}\sum_M \Braket{U|\Phi_M}   \left( \Braket{\Phi_M}_{Y_{IJ}}^{\left(+\right)} - \Braket{\Phi_M}_1 \right),
\end{eqnarray}
\end{widetext}
where $\Braket{\quad}_{Y_{IJ}}^{\left(+\right)}$ denotes taking linear average over all possible atomic configuration satisfying $Y_{IJ}\ge \Braket{Y_{IJ}}_1$. 
Eq.~\eqref{eq:y} directly means that we can construct PS to characterize temperature dependence of selected pair probability $Y_{IJ}$, where its structure is given by 
$\left\{ \Braket{\Phi_1}_{Y_{IJ}}^{\left(+\right)},\cdots, \Braket{\Phi_f}_{Y_{IJ}}^{\left(+\right)} \right\}$ for $f$-dimensional configuration space considered, and structure of SQS is 
given by $\left\{\Braket{\Phi_1}_1,\cdots,\Braket{\Phi_f}_1\right\}$. 
Note that defiition of the above pair probabily for unlike-atom pair includes permutation of constituent pair probability $y_{IJ}$ and $y_{JI}$, namely,
\begin{eqnarray}
Y_{IJ} = y_{IJ} + y_{JI},
\end{eqnarray}
where constituent pair probabilities satisfy the following summation for composition of $I$, $c_I$:
\begin{eqnarray}
\sum_{J} y_{IJ} = c_I.
\end{eqnarray}

%\item
We here consider ordering tendency in terms of 1st nearest-neighbor (1NN) pair probability for seven fcc-based equiatomic ternary alloys of CrCoNi, CrFeNi, CrNiMn, CoNiMn, FeNiMn, AgAuCu and AgPdRh: The reasons for choosing these alloys are (i) Ni-based ternarly alloys are all subsystems of HEAs whose short-range order is considered as key role to characterize its extreme mechanical properties, and (ii) compared to binary alloys, very little has been known for short-range order even qualitatively, where short-range ordering tendency for Ag-based two alloys, CrCoNi and CrFeNi are qualitatively available for previous experimental and/or theoretical studies. 

%\item
In the present study, PSs along possible six pairs (A-A, A-B, B-B, A-C, B-C and C-C) and SQS are constructed for supercell of 480-atom ($4\times5\times6$ expantion of convensional unit cell) based on Monte Carlo simulation to minimize Euclidean distance between multisite correlation functions for practically-constructed configuration and that for ideal values,\cite{em1} where we consider up to 6NN pair, and all triplets and quartets consisting of up to 4NN pairs that can well characterize ordering tendency for fcc-based binary alloys. We emphasize again that structure of PSs and SQS are constructed only for geometric information of underlying lattice, which are all common for the seven ternary alloys.

%\item
The constructed PSs and SQS are applied to density functional theory (DFT) calculation to obtain total energy, which is perfomed by the VASP\cite{vasp} code using the projector-augmented wave method,\cite{paw} with the exchange-correlation functional treated within the generalized-gradient approximation of Perdew-Burke- Ernzerhof (GGA-PBE).\cite{pbe} The plane wave cutoff of 360 eV is used. Structural optimization is performed until the residual forces less than 0.005 eV/\AA.

%\end{itemize}

\section{Results and Discussions}
\begin{figure}[h]
\begin{center}
\includegraphics[width=0.93\linewidth]
{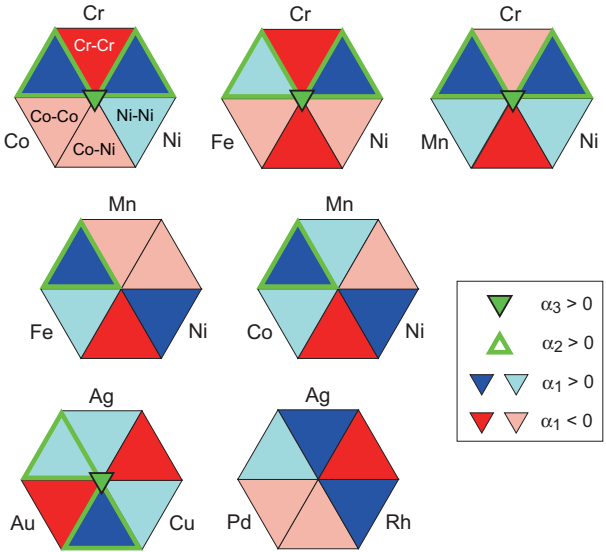}
\caption{Color plot of sign and magnitude of $\alpha_3$, $\alpha_2$ and $\alpha_1$. For $\alpha_1$, dark blue and dark red color triangles denotes pairts with the top three highest absolute value of $\alpha_1$.  }
\label{fig:a}
\end{center}
\end{figure}

%\begin{itemize}
%\item
For multicomponent system, information about whether a chosen unlike-atom pair is preferred or disfavored w.r.t. random state is not sufficient to characterize whether the system undergoes ordered structure or phase separation at low temperature, which is in contrast to binary systems. This is simply because for multicomponent systems, there are multiple (for ternary, three) different kinds of unlike-atom pair.
%\item
Therefore, we here introduce two type of quantity to measure the ordering (or separating) tendency as follows:
\begin{eqnarray}
\alpha_3 &=& -\frac{1}{2} \sum_{I\neq J} U_{IJ} + \sum_J U_{JJ}, \nonumber \\
\alpha_2^{\left(JK\right)} &=& -U_{JK} + \left(U_{JJ} + U_{KK}\right), \nonumber \\
\alpha_1^{\left(JK\right)}  &=& -U_{JK}, \nonumber \\
&&\left(I, J, K = \textrm{A, B, C} \right),
\end{eqnarray}
where $U_{JK}$ denotes potential energy of PS for $JK$ pair, given in Eq.~\eqref{eq:y}. 
Note that here and hereinafter, for simplicity PS energy and its correlation functins are always measured from SQS energy and correlation functions. 
%\item
From the definition, we can see that (i) $\alpha_3$ denotes the measure of whole preference (disfavor) for three unlike-atom pairs
w.r.t. other three like-atom pairs when $\alpha_3$ exhibit positive (negative) sign, 
(ii) $\alpha_2$ denotes the measure of preference (disfavor) for a selected unlike-atom pair w.r.t. corresponding two like-atom pairs when 
$\alpha_2$ exhibits positive (negative) sign, (iii) $\alpha_1$ denotes the measure of preference (disfavor) for a selected unlike-atom or like-atom pair 
w.r.t. random states with its positive (negative) sign, and (iv) for ideally random states, $\alpha_3=\alpha_2=\alpha_1=0$.
%\item
Therefore, $\alpha_3$ and $\alpha_2$ respectively represent the measure of overall preference (disfavor) of unlike-atom pair(s) for ternary system and 
three subsystems, i.e., constituent binary systems. 

%\item
Figure~\ref{fig:a} shows the color plot of sign and magnitude of $\alpha_3$, $\alpha_2$ and $\alpha_1$ for the present seven ternary alloys. 
We can clearly see that ordering (or separating) tendency for a chosen pair in terms of $\alpha_1$ can strongly depend on combination of the rest element: For instance, Ni-Ni like-atom pair is strongly preferred for FeMnNi and CoMnNi, while it is disfavored for CrFeNi alloy. The fact that ordering tendency as ternary (i.e., $\alpha_3 > 0$) does not always holds for its subsystems, e.g., for CrCoNi, $\alpha_3 > 0$ while for Co-Ni subsystem, $\alpha_2 < 0$. Moreover, we can clearly see for Fe-Ni subsystem in CrFeNi and Co-Mn subsystem in CoMnNi alloys the counterbalance breaking of like- and unlike-atom pair that is not broken for binary system: Fe-Ni subsystem exhibit the same negative sign for Fe-Fe, Fe-Ni and Ni-Ni pair, and Co-Mn exhibit the same positive sign for Co-Co, Mn-Mn and Co-Mn pairs.

\begin{figure}[h]
\begin{center}
\includegraphics[width=0.97\linewidth]
{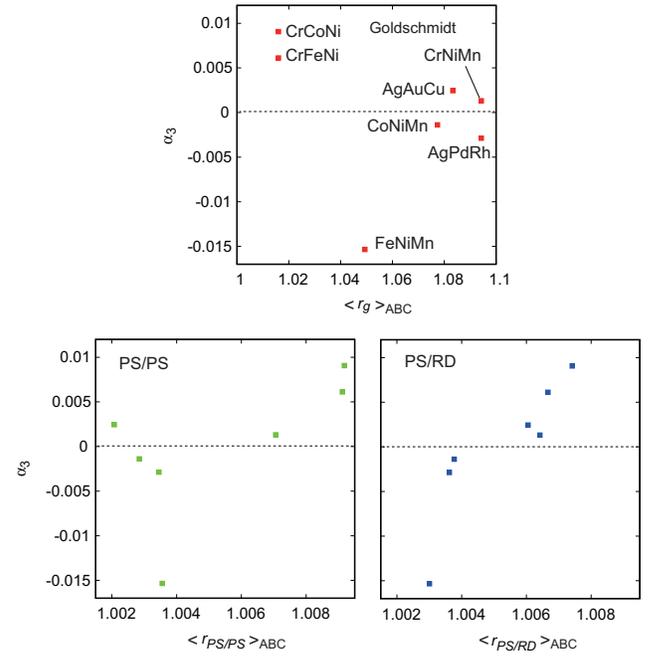}
\caption{Ternary ordering tendency in terms of averaged ratio of atomic radius under three different definitions.}
\label{fig:a3}
\end{center}
\end{figure}

\begin{figure}[h]
\begin{center}
\includegraphics[width=0.97\linewidth]
{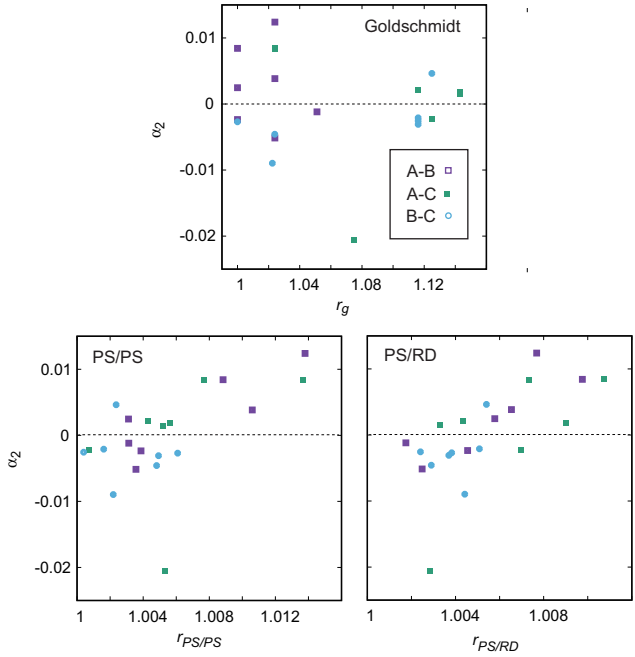}
\caption{Binary-subsystem ordering tendency in terms of ratio of atomic radius under three different definitions.}
\label{fig:a2}
\end{center}
\end{figure}

%\item
In order to characterize the above ordering tendencies in terms of underlying geometry, we also introduce three kinds of atomic radius ratio as follows:
\begin{eqnarray}
r_g^{\left(JK\right)} &=& \left[ \frac{R_{g}^{\left(J\right)}}{R_{g}^{\left(K\right)}} \right]_1 \nonumber \\
r_{PS/PS}^{\left(JK\right)} &=& \left[ \frac{R_{PS}^{\left(J\right)}}{R_{PS}^{\left(K\right)}} \right]_1 \nonumber \\
r_{PS/RD}^{\left(JK\right)} &=& \frac{1}{2}\left(  \left[ \frac{R_{PS}^{\left(J\right)}}{R_{RD}} \right]_1 +  \left[ \frac{R_{PS}^{\left(K\right)}}{R_{RD}} \right]_1 \right), 
\end{eqnarray}
where $\left[\quad\right]_1$ denotes that internal numerator and enumerator can be reversed so that resultant value of fraction should always be greater or equal to 1, and $R_g^{\left(K\right)}$, $R_{PS}^{\left(K\right)}$ and $R_{RD}$ respectively denotes Goldschmidt atomic radius\cite{gold} for element $K$, effective atomic radius of $K$ obtained from projection state, and effective atomic radius for random states (i.e., from SQS structure). Note that since PS contains multiple elements other than considered $K$, we define $R_{PS}^{\left(K\right)}$ from the volume of the following conditions:
\begin{eqnarray}
\label{eq:vjk}
V_{JK} = V_{SQS} + C\cdot \left(V_{PS}^{\left(JK\right)} - V_{SQS}\right), 
\end{eqnarray}
where constant $C$ is determined so that 
\begin{eqnarray}
\label{eq:ykk}
\left |Y_{KK}\left( C\cdot\left\{ \Braket{\Phi_1}_{Y_{IJ}}^{\left(+\right)},\cdots, \Braket{\Phi_f}_{Y_{IJ}}^{\left(+\right)} \right\} \right) -1 \right| = \textrm{min}.
\end{eqnarray}
With these prerations, $R_{PS}^{\left(K\right)}$ can be obtained through
\begin{eqnarray}
\label{eq:rps}
R_{PS}^{\left(K\right)} \propto \left\{ \left(\sum_{I\neq J} V_{IJ}/2\right) - V_{LM}\left(L\neq K, M\neq K\right)  \right\}^{1/3}
\end{eqnarray}
for system with $\alpha_3 > 0$, and 
\begin{eqnarray}
\label{eq:rps2}
R_{PS}^{\left(K\right)} \propto \left\{ V_{KK}/2 \right\}^{1/3}
\end{eqnarray}
for system with $\alpha_3 < 0$.
The reason why we do not employ sign of $\alpha_2$ or $\alpha_1$ to determine effective atomic radius is that (i) there can be multiple values for atomic radius for a chosen element of a given system, and (ii) algebraic equations for atomic radius can be linear dependent  (e.g., CrMnNi system: there is four preference pairs) with each other when we focus on the preference of ordering tendency of subsystems or cannot be solved (e.g., CrFeNi system: there is two preference pairs). 
%\item
Eqs.~\eqref{eq:vjk} and~\eqref{eq:ykk} means that (i) since projection state in Eq.~\eqref{eq:y} corresponds a specially fluctuated structure (i.e., covariance fluctuation) from random state along considered pair (similarly to binary system), we define effective atomic radius from information about such fluctuated structure, (ii) it has uncertainity to multiply scalar constant $C$ for projection state since only the direction of fluctuation is essential to determine pair probability (invariant to the choice of $C$), while resultant effective volume (or atomic radius) depends on $C$, and (iii) therefore, we determine the magnitude of fluctuation $C$ so that the fluctuated structure has pair probability corresponding to the value of correlation function for considered pure element (Eq.~\eqref{eq:ykk}). 
%\item
Eqs.~\eqref{eq:rps} and~\eqref{eq:rps2} means that effective atomic radius are determined from information about volume of fluctuated structure with prefered direction (i.e., for $\alpha > 0$, fluctuation along unlike-atom pair, and $\alpha < 0$, along like-atom pair). 

%\item
Figure~\ref{fig:a3} shows the resultant ternary atomic ordering tendency, $\alpha_3$, in terms of the above three different kinds of definition of atomic radius ratio. $\Braket{\quad}_{\textrm{ABC}}$ means taking linear average over possible set of pairs, e.g., for Goldschmidt radius, 
\begin{eqnarray}
\Braket{r_g}_{\textrm{ABC}} = \frac{1}{3}\left( r_g^{\left(\textrm{AB}\right)} + r_g^{\left(\textrm{AC}\right)} + r_g^{\left(\textrm{BC}\right)} \right).
\end{eqnarray}
%\item
We can clearly see that Goldshmidt atomic radius ratio does not exhibit effective correlation with ordering tendency: systems with larger positive value of $\alpha_3$ does not have larger value of $\Braket{r_g}_{\textrm{ABC}}$. The correlation is most strong when we take $\Braket{r_{PS/RD}}_{\textrm{ABC}}$ compared with $\Braket{r_{PS/PS}}_{\textrm{ABC}}$, and when we decompose these tendency into subsystems (see Fig.~\ref{fig:a2}), the same correlation relationships holds true.  These strongly indicate that for multicomponent disordered alloys, geometric effects on ordering tendency in terms of strain due to difference in atomic radius should be naturally considered as difference in atomic radius between effective constituent radius in covariance-fluctuated strucre and that for random states. 
%\end{itemize}

\section{Conclusions}
Based on specially fluctuated structure and ideally random structure, we here show that introduced definition of atomic radius can reasonablly characterize SRO tendency for multicomponent alloys not only for whole (ternary) system as well as their constituent subsystems, while conventional definition of atomic radius fails to explain such SRO tendency. The present findings strongly indicate the significant role of geometry in underlying lattice on SRO for multicomponent alloys, where quantitative formulation of SRO parameter in terms of the present geometric factors should be performed in our future study.

\section{Acknowledgement}
This work was supported by Grant-in-Aids for Scientific Research on Innovative Areas on High Entropy Alloys through the grant number JP18H05453 and a Grant-in-Aid for Scientific Research (16K06704) from the MEXT of Japan, Research Grant from Hitachi Metals$\cdot$Materials Science Foundation, and Advanced Low Carbon Technology Research and Development Program of the Japan Science and Technology Agency (JST).

\end{document}